\documentclass[12pt]{article}
\usepackage{pdproc,epsfig} 

\begin{document}

\renewcommand{\thefootnote}{\alph{footnote}}
\begin{flushright}
DFPD-01/EP38
\end{flushright}
\title{
 SPECTROSCOPY OF LOW ENERGY NEUTRINOS \\
 FROM THE SUN} 
\author{CARLO BROGGINI}
\address{Istituto Nazionale di Fisica Nucleare, Sezione di Padova,
 Via Marzolo 8, 35131 Padova, Italy\\E-mail: broggini@pd.infn.it} 
\abstract{Two methods are discussed for the solar neutrino 
spectroscopy in the sub-MeV region: absorption in a loaded liquid
scintillator and elastic scattering in a TPC.
The different neutrino oscillation solutions predict a strong
effect in this energy region where the largest fraction
($\sim$98$\%$)
of solar neutrinos lies.
Both projects have 
reached the stage where they have to prove their
capability to attain a background low enough for solar neutrino 
detection.}
\normalsize\baselineskip=15pt
\section{ Introduction}
The largest fraction of the calculated standard 
model solar neutrino flux lies
below 1 $MeV$\cite{ba2000}. In particular the $pp$ neutrinos
are the most abundant source of solar neutrinos (about 91$\%$ of 
the total flux) and they are calculated to 1$\%$ accuracy,
whereas 
the $^{7}Be$ neutrinos give a contribution of about 7$\%$ to the
total flux and they are calculated to 10$\%$ uncertainty.

The different neutrino oscillation solutions predict 
a strong influence on the low energy $\nu_{e}$
from the Sun\cite{ba2001}. A measurement of the solar
neutrino spectrum
in the sub-MeV region might be the only possibility 
of having a precise determination of the mass difference and
the mixing angle relevant to the solar neutrino 
phenomena. This is particularly true if 
Kamland, the long-baseline reactor experiment, will not see
anti-neutrino oscillations in the $\delta$$m^{2}$
region down to $\sim$ 10$^{-5}$ $eV^{2}$.
At the same time the measurement of the $pp$ neutrinos 
would be an important test of the stellar evolution theory.

All the running experiments sensitive to the sub-MeV neutrinos
from the Sun are radiochemical (Homestake, Gallex/GNO, Sage)
and cannot do any neutrino spectroscopy. BOREXINO will start
taking data next year and it is so far the only 
real time experiment
which can detect low energy neutrinos by measuring the
energy of the recoiling electron in the reaction
${\nu}e^{-} \rightarrow {\nu}e^{-}$. The 250 $keV$  
threshold on the electron kinetic energy allows 
only the detection of the $^{7}Be$ neutrinos.

In this paper I will describe two different approaches 
which are studied for the spectroscopy of both the $pp$ and
$^{7}Be$ neutrinos. 
First I will discuss the neutrino 
detection through the  
${\nu}_{e}+(A,Z) \rightarrow {(A,Z+1)^{*}}+e^{-}$ 
interaction in a liquid scintillator detector and then,
in a more detailed way, the detection through the 
${\nu}e^{-} \rightarrow {\nu}e^{-}$ 
elastic scattering  
in a gas filled Time Projection Chamber.
\vfill
\eject
\section{ Loaded liquid scintillator: LENS}
In 1997 R.Raghavan proposed\cite{ra97} to detect the low energy 
$\nu_{e}$ from the Sun through the charged current interaction
$\nu_{e}+{^{176}Yb}
\rightarrow {^{176}Lu^{*}}+e^{-}$.
The neutrino energy is simply given by the sum of the electron 
kinetic energy and of the $Q$ value of the reaction (301 $keV$).
The delayed coincidence between the electron from the
$\nu_{e}$ interaction and the gamma ray from the de-excitation
of $^{176}Lu^{*}$ ($E_{\gamma}=71.5 keV, \tau=50 ns$)    
gives the $\nu_{e}$ tag.
In addition $^{176}Yb$ has a reasonable isotopic abundance
(12.8$\%$) and it is a stable nucleus against single $\beta$
decay (it is an even-even $\beta\beta$ candidate nucleus).

Since the proposal of Raghavan a strong research program has been 
undertaken by the LENS Collaboration to develop a liquid scintillator
with high and stable light yield at high $Yb$
content. An important technical success has been achieved at the beginning 
of this year with the production of an aromatic scintillation solvent 
loaded with organo-metallic $Yb$ compounds with the following 
features\cite{ra01}: light yield of 8400 photons/$MeV$
with 10$\%$ $Yb$ loading, attenuation 
length of a few meters and stability in time during about 1 year.
A modular detector could be filled with this scintillator to
have, with 20 $tons$ of Ytterbium, a rate of 180 $pp$ and 143 
$^{7}Be$
events/year.   

The background studies give a less promising perspective. 
It is true that, thanks to the coincidence, it is possible to release
the requirements on the contaminants which give 
random coincidences of uncorrelated
background (by about 6 orders of magnitude as compared to BOREXINO),
however severe sources of correlated background have been 
identified. For correlated background we mean all the decay schemes
which give two signals correlated in time and space as the 
$\nu_{e}$ tag, such as 
$^{231}Th$ ($^{235}U$ chain), $^{169}Yb$ (from
$n$ capture on $^{168}Yb$), $^{176}Lu$ (a rare earth difficult
to remove from Ytterbium). 

Shortly, the technical problems related to the
production of the $Yb$ loaded scintillator seem to be solved, but
the possibility of reaching an acceptable background has
still to be demonstrated. This is one of the main items for the
LENS low background facility which is almost ready in
the Gran Sasso underground laboratory.

This spring Raghavan succeeded in producing an Indium
loaded liquid scintillator\cite{ra02}, using the same
recipe as for Ytterbium and obtaining similar results.
Since 25 years $^{115}In$ is the dream of the physicists
aiming at the low energy solar neutrino detection.
The detection scheme is the following:
$\nu_{e}+^{115}In \rightarrow ^{115}Sn^{*}+e^{-}$.
The de-excitation of $^{115}Sn^{*}$ ($\tau$=4.76 $\mu$s)
gives rise to a 115.6 $keV$ electron (or $\gamma$ in 4$\%$
of the decays) and to a 497.3 $keV$ $\gamma$. 

Indium is so appealing as a neutrino detector medium for the following
reasons: low $Q$ value (118 $keV$), delayed coincidence 
with strong signature and of
relatively high total energy (613 $keV$), high isotopic abundance
(95.7$\%$).

The most serious problem, 
which till now prevented the use of Indium as a 
neutrino detector medium, 
is the beta decay of $^{115}In$ itself
($\tau=6 \cdot 10^{14}$ years, end-point energy=495 $keV$, 
activity: 0.25 $Bq/gr$).
The relatively high light output of the new scintillator
might give an energy resolution good enough to
efficiently distinguish the $^{115}Sn^{*}$ 
de-excitation
(which is the delayed coincidence to tag the $\nu_{e}$
interaction) from the beta decay with the emission of a
Bremsstrahlung photon of $^{115}In$ (this is the most 
severe background source due to the Indium
radioactivity\cite{ra02}). As a matter of fact, the
former event has a total energy of 613 $keV$ 
whereas the latter one can have, at most, a 495 $keV$
energy. 
In addition to energy resolution, the granularity 
is the other key feature required by an Indium detector 
to fully exploit the strong signature of the $\nu_{e}$ 
absorption.  Both the points are now studied to check
if a background 
could be achieved  
low enough for solar neutrino detection.
\section{Solar TPC}
The great advantage of detecting the solar neutrinos by 
measuring both the energy and the direction of
the recoiling electron in the 
${\nu}+e^{-} \rightarrow {\nu}+e^{-}$
elastic scattering is shown by
the Kamiokande and SuperKamiokande 
experiments with the high energy $^{8}B$ neutrinos.
As a matter of fact, the water $\check{C}$erenkov detectors would have
not been able to identify any solar neutrino signal
without the reconstruction of the electron direction.
Actually also the $\mu$ and $\tau$ neutrinos scatter 
electrons, but with a lower cross section ($1/7$-$1/6$ of the
$\nu_{e}$ one). 

The measurement of both the 
electron energy and direction allows for:
\begin{itemize}
\item the solar neutrino astronomy, with an unambiguous signal 
identification,
\item the spectroscopy of the neutrinos from the Sun
(the $\nu_{e}$ energy is reconstructed from the electron
energy and direction),
\item the on-line measurement of the background, 
\item a higher signal to noise ratio.
\end{itemize}

The threshold on the electron energy has to be kept low,
100 $keV$, to be sensitive to both the $pp$ and $^{7}Be$
neutrinos.
At low energy the detector
medium has to be a gas because 
the recoiling 
electron track must have a minimum length ($\sim$3 $cm$) in order 
to be reconstructed. The electrons of the gas are the targets 
for the $\nu e$ scattering.

The HELLAZ TPC
has been firstly discussed in\cite{hel01} and a Letter of Intent has
been submitted last January. It concerns a 
2000 $m^{3}$ chamber filled with 7 tons of $He+CH_{4}$ (95/5).
Such a gas density (3.5 $gr/l$) can be reached with different 
combination of pressure and temperature: 20$bar$/300$K$, 
10$bar$/140$K$, 5$bar$/77$K$. A recent review of the project 
is given in\cite{bon01}.

In this paper I will discuss, instead, the project 
of a solar TPC filled with $CF_{4}$ at 1 $bar$ pressure. 
$CF_{4}$ has at the same time
a low atomic number
and a high density (3.7 $gr/l$
at 1 $bar$ and room temperature).
The low atomic number minimizes the multiple
scattering, allowing for a good track reconstruction,
whereas the high density maximizes the number of target electrons.
I will found my discussion 
on the results of the MUNU 
experiment\cite{bro92}\cite{am97}\cite{av01}.

MUNU was designed to study the 
$\overline{\nu}_{e}e^{-} \rightarrow \overline{\nu}_{e}e^{-}$ 
scattering with the antineutrinos from a nuclear
reactor.
\begin{figure}[htb]
\begin{center}
\epsfbox{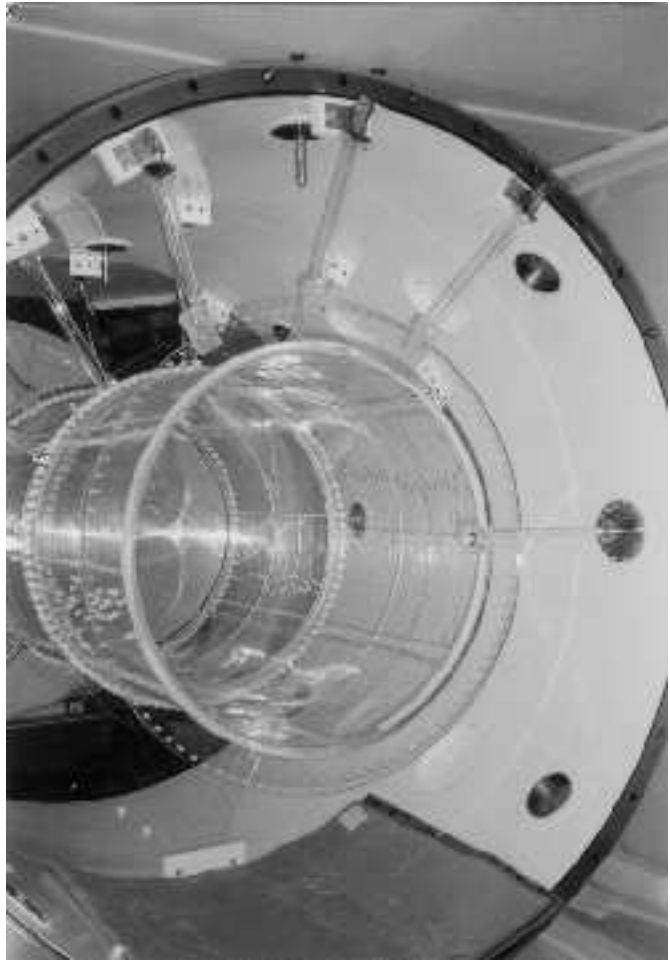}
\caption{The TPC mounted inside the anti-Compton without the anode 
plane.}
\label{fig:opend}
\end{center}
\end{figure}
\begin{figure}[t]
\begin{center}
\epsfxsize=7.5cm
\epsfbox{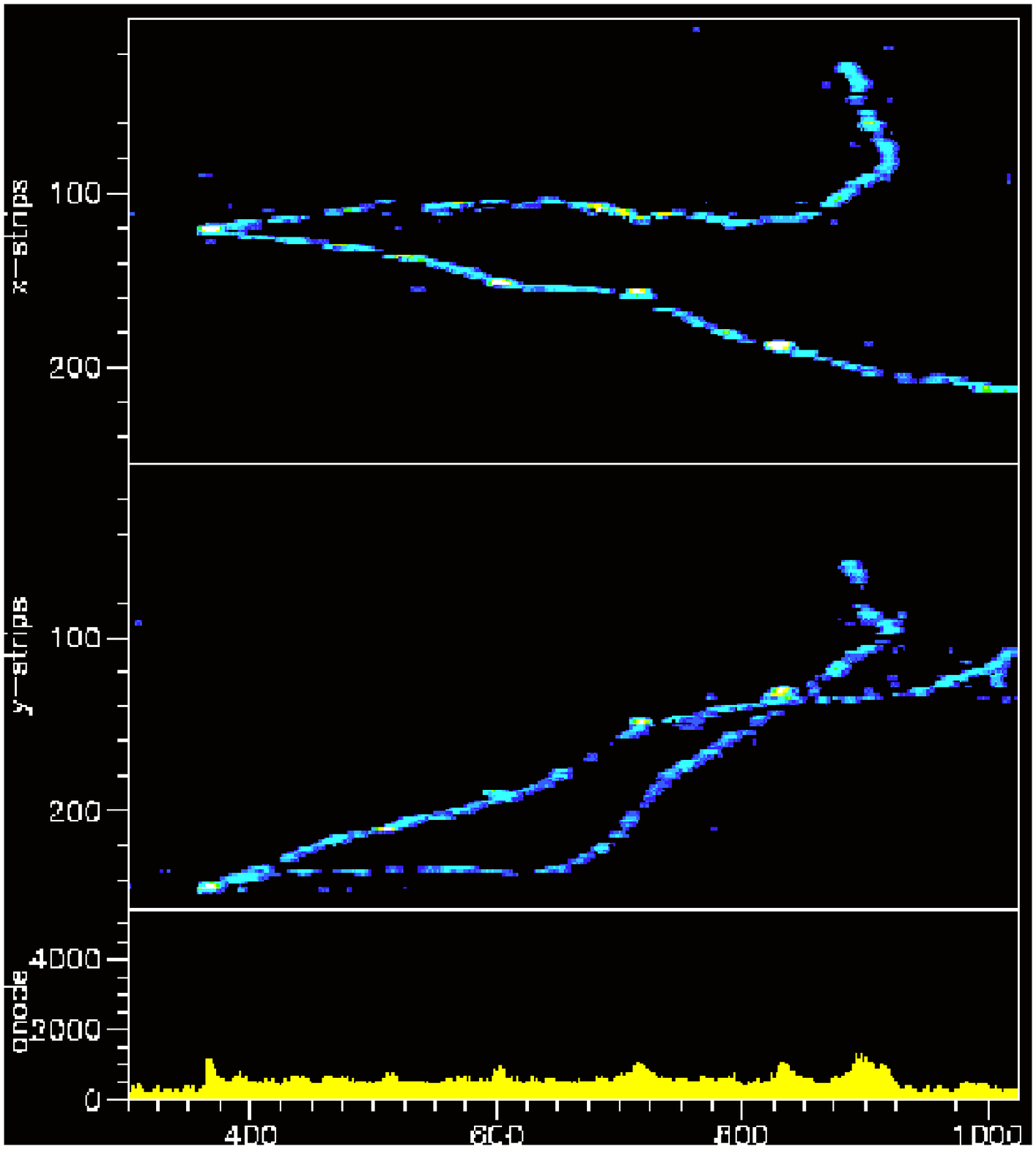}
\epsfxsize=7.5cm
\epsfbox{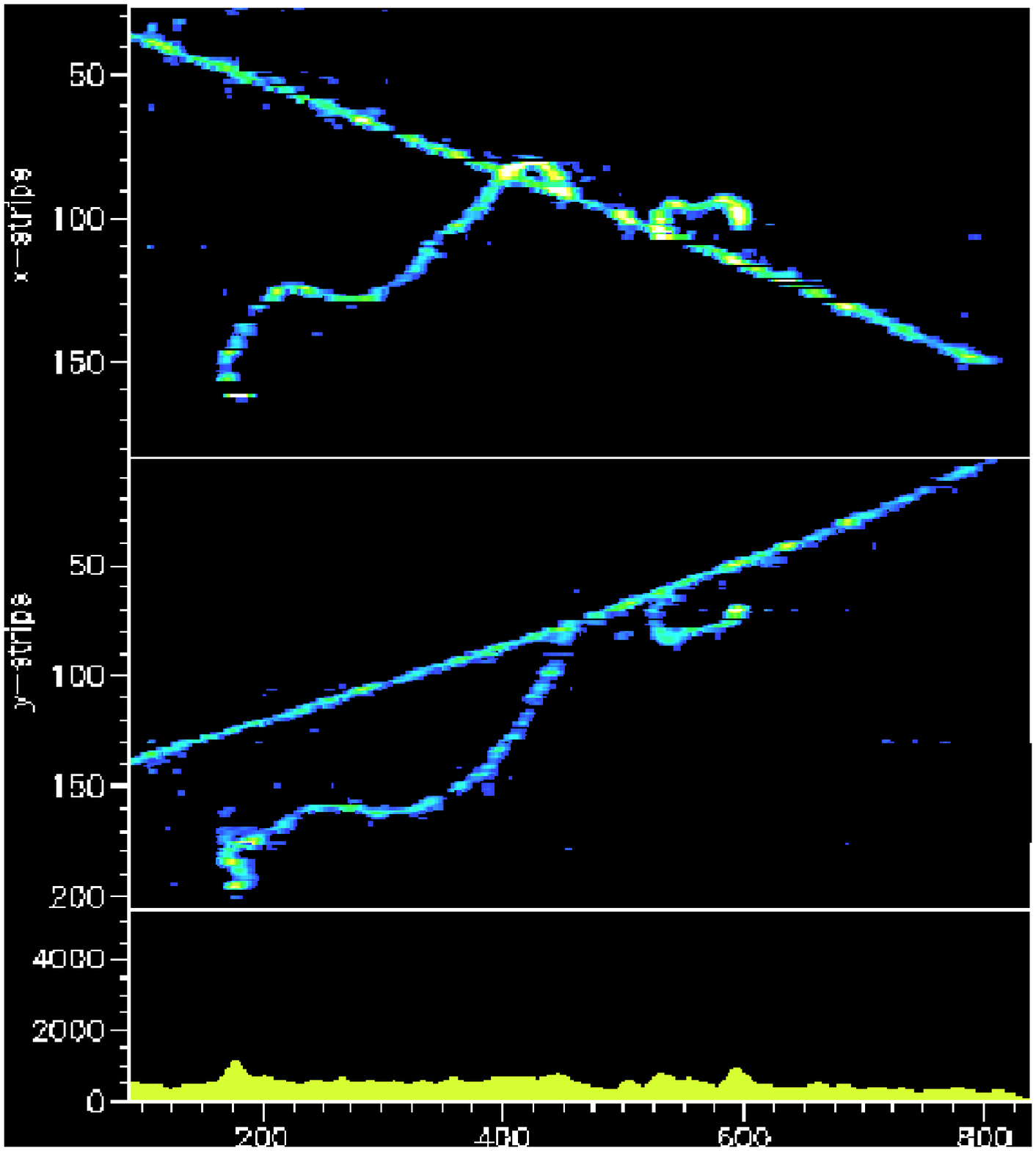}
\caption{An $e^{+}e^{-}$ event and a 
 cosmic muon going through the TPC with the production
of 2 deltas: 
vertical projection, horizontal projection
and energy deposition as function of time. The binning is 3.5 $mm$ for $x$ and
$y$ and 80 $ns$ for the time (i.e. 1.7 $mm$ for $z$).}
\label{fig:carlo_3}
\end{center}
\end{figure}
\begin{figure}[htb]
\begin{center}
\epsfxsize=7.5cm
\epsfbox{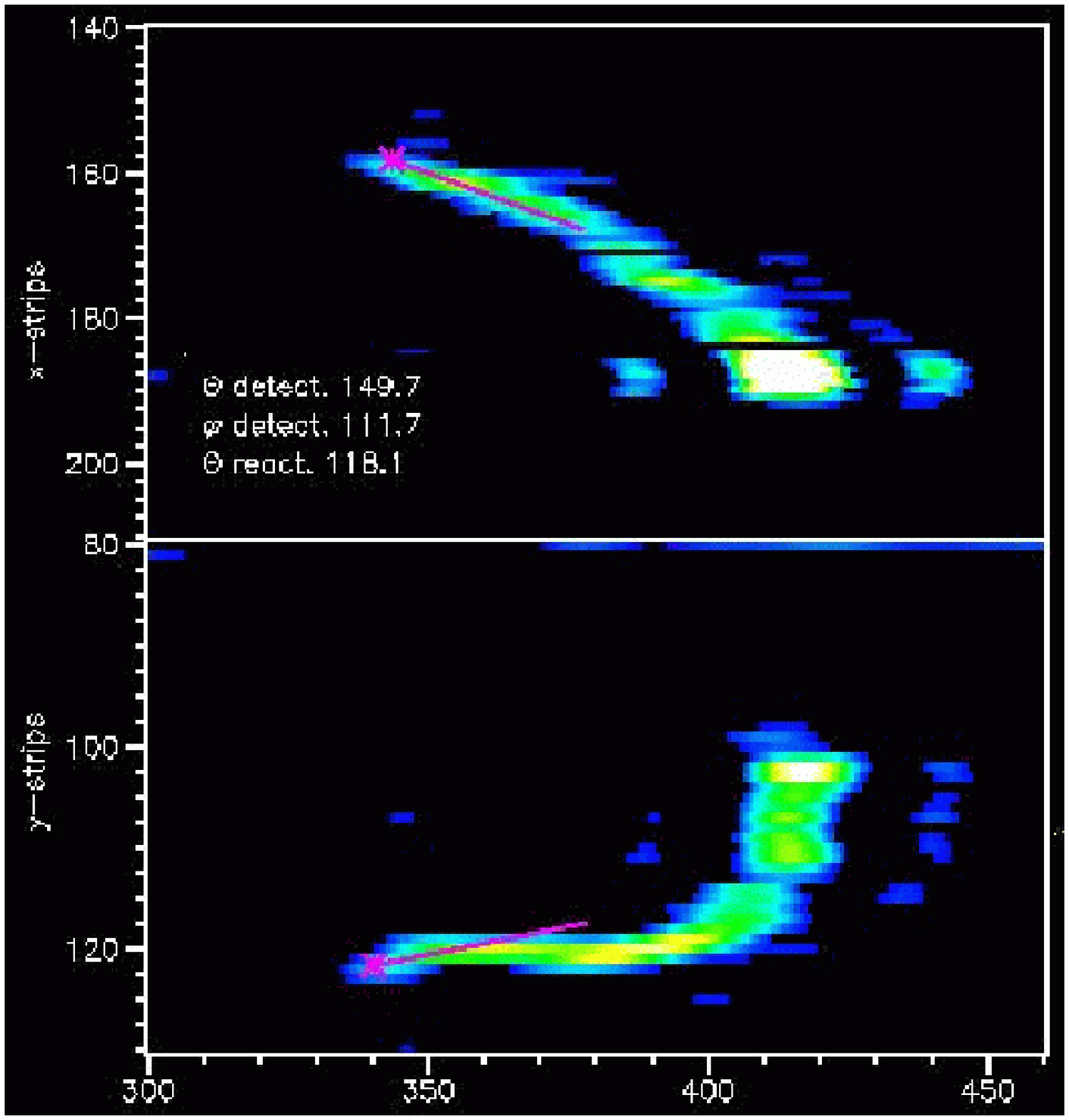}
\epsfxsize=7.5cm
\epsfbox{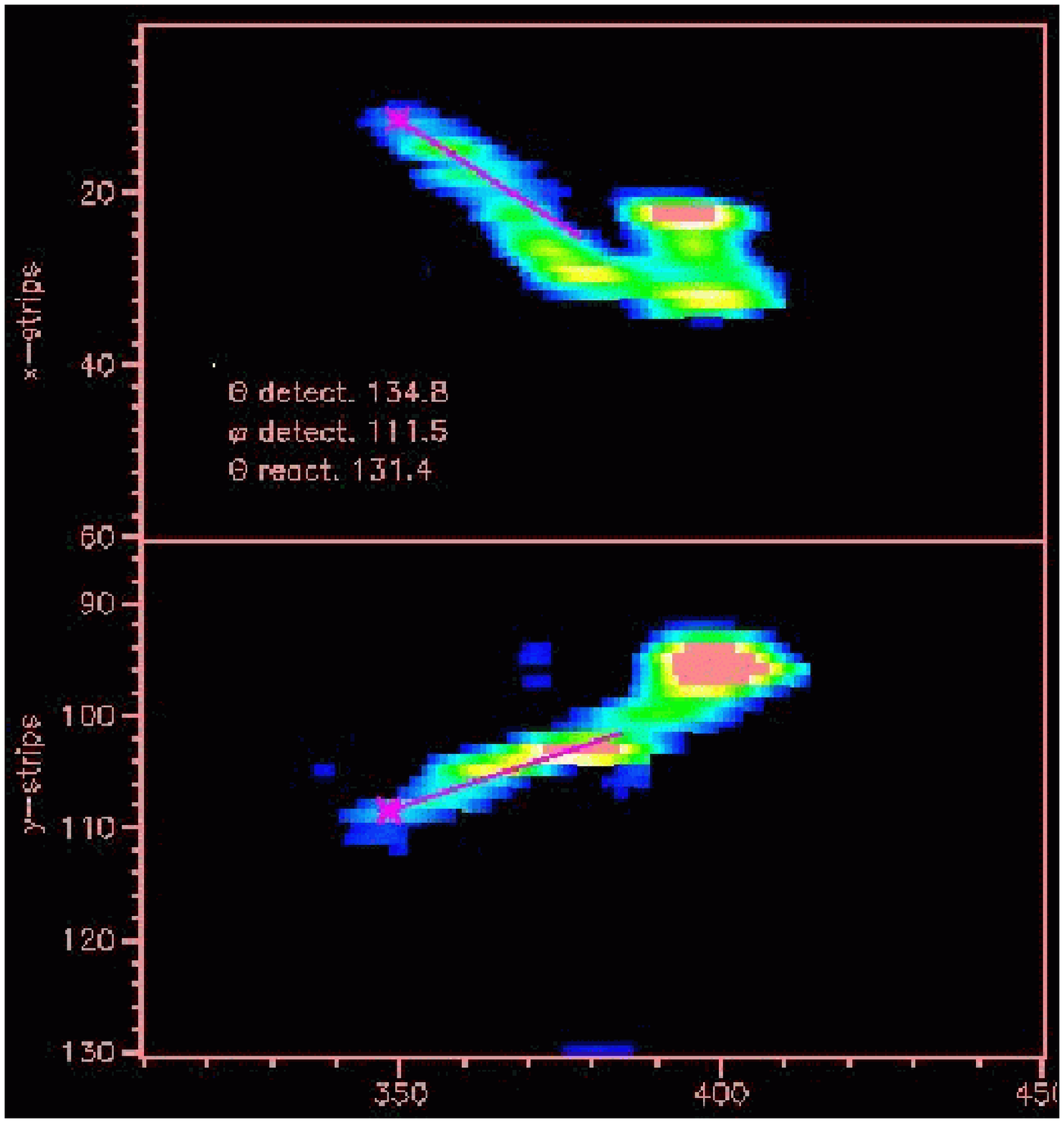}
\caption{Low energy electrons of 590 $keV$ and 460 $keV$ energy
 (18 $cm$ and 12 $cm$ range, respectively).}
\label{fig:carlo_5}
\end{center}
\end{figure}
The detector is running in the Bugey laboratory,
at a distance of 18$m$ from the core of a 2800$MWth$ reactor.
It consists of a 1 $m^{3}$ TPC immersed into 10 $m^{3}$ of liquid
scintillator working as anti-Compton
(Fig.\ref{fig:opend}).
Since the anti-neutrino 
event rate is low it has been necessary to 
minimize all possible background sources and to construct
each part  
of the detector from selected low radioactivity 
material.

The central detector is an acrylic vessel TPC,
a cylinder 
of 90 $cm$ inner diameter and 162 $cm$ long, filled with $CF_{4}$
at 3 $bar$ pressure.
Acrylic is selected as construction material because of its 
low intrinsic radioactivity\cite{bro93}.

The acrylic TPC is mounted inside a stainless steel tank 
(3.8 $m$ long and 2 $m$ in diameter)
filled with 
liquid scintillator.
The liquid scintillator, 50 $cm$ thick, 
serves to veto the cosmic muons
and as anti-Compton detector. 
It is viewed by 48 hemispherical photomultipliers, 
24 on each lid, of
20 $cm$ diameter and made with low activity glass. 

Outside the steel vessel
there are a 8 $cm$ thick boron loaded polyethylene shielding 
and a 15 $cm$ thick lead shielding to absorb neutrons and gamma
rays entering 
the detector from outside.

\begin{figure}[htb]
\begin{center}
\epsfxsize=7.5cm
\epsfbox{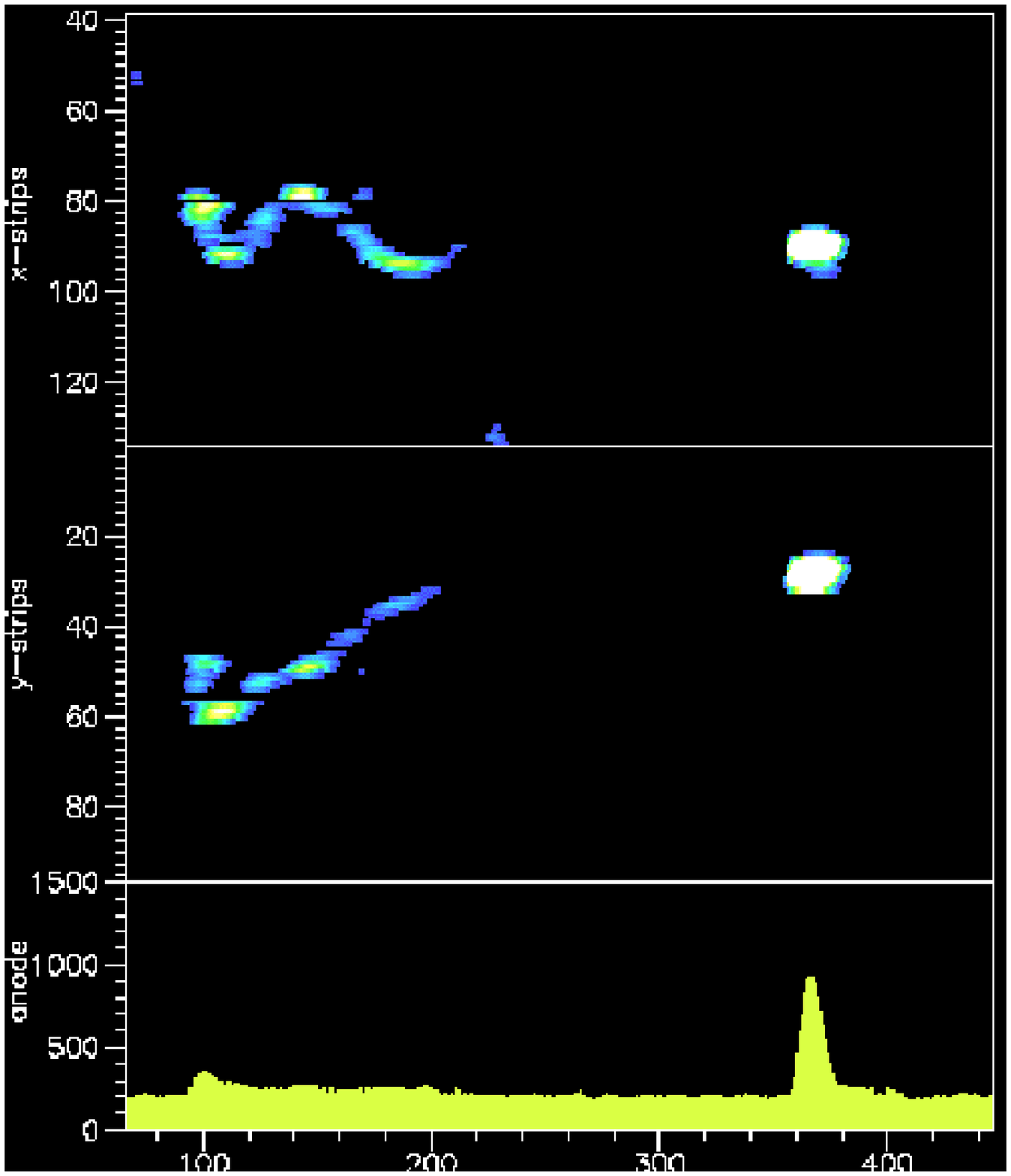}
\caption{An electron-$\alpha$ event due to a $Bi-Po$
decay. The $\alpha$ can be easily identified because of the short track
and the large energy deposition. \label{fig:carlo_8}}
\end{center}
\end{figure}
The MUNU detector is the first one 
able to see the event peak due to the $\overline\nu_{e}$ from 
a nuclear reactor and able to do 
the neutrino spectroscopy in the region below 1 $MeV$.
As a matter of fact, it does exactly the same as what a 
solar neutrino TPC should do
(the only difference is the flux
at the detector site: the neutrino flux from the Sun 
is more than two orders of magnitude lower than the anti-neutrino flux
from the reactor). 
As a consequence, it can be 
regarded as a low background
prototype of a solar TPC.
Clearly the detector is not optimized to work at 1 $bar$, many of 
the technical solutions we selected cannot be adopted in a much bigger 
detector and, finally, it is running in a place far different from a 
'silent' underground laboratory.
However the results it has provided on track 
reconstruction, energy resolution and
background suppression can be regarded as a sound starting point
for a solar neutrino TPC project.

Fig. \ref{fig:carlo_3} shows an $e^{+}e^{-}$ event and a cosmic muon, 
whereas in
fig. \ref{fig:carlo_5}
there are two 
low energy  
electron events.
The electron energy and angular resolution are such that 
it has been possible to reconstruct the $^{54}Mn$
photo-peak ($\gamma$ energy=835 $keV$) with a $\sigma$
of 220 $keV$ by measuring the energy
and the direction of the Compton electrons inside the 
TPC\cite{av01}.
This is an encouraging results because it shows that the $^{7}Be$
neutrino peak (862 $keV$) could be reconstructed in a TPC filled
with $CF_{4}$ even at 3 $bar$ pressure and with a 300 $keV$
threshold on the kinetic energy of the recoiling electron.

The background requires a more detailed discussion.
The electron and 
$\alpha$ particle rates 
were at the beginning much higher than the predicted ones
because of
the $^{222}Rn$ emitted by the oxysorb through which the gas is 
going to remove Oxygen and water.

As a matter of fact, the Radon
signature can be easily seen in our detector (fig. \ref{fig:carlo_8}):
an electron from the $\beta$ decay of
$^{214}Bi$ (end-point: 3.26 $MeV$) followed by the $\alpha$
decay of $^{214}Po$ ($\alpha$ energy: 7.8 $MeV$, half-life: 164
$\mu$s).

After having replaced the oxysorb 
and changed the TPC cathode, where the Radon daughters 
were collected,
we reached an event rate over the 4$\pi$ solid angle of 350 cpd 
(counts per day) and 50 cpd for electron kinetic energy 
above 300 $keV$ and 800 $keV$, 
respectively. We remark that such rates, when divided by the 11.1 $kg$
of the $CF_{4}$ mass, are very similar to the cpd/$kg$ measured in
Gran Sasso by low background Germanium detectors surrounded by 
Copper, Lead and with an anti-Radon shielding. 

The active shielding,
the scintillation light of $CF_{4}$ itself
and the the possibility of defining a fiducial volume
have been particularly 
helpful in reducing the background of the MUNU detector. 
I think that they are key features of the detector
which must also be kept in the solar TPC.

\begin{figure}[htb]
\begin{center}
\epsfxsize=8.5cm
\epsfbox{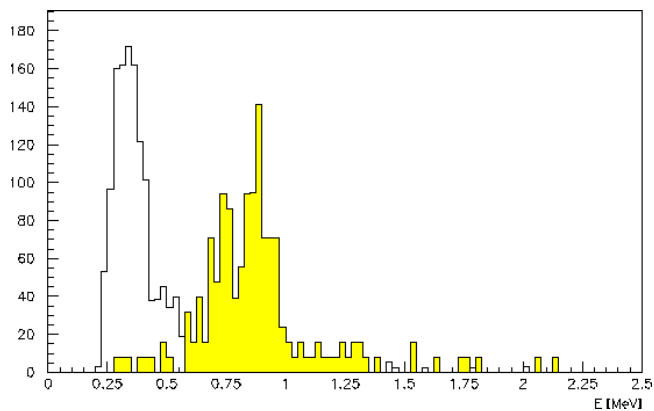}
\epsfxsize=8.5cm
\end{center}
\caption{Simulation of the $pp$  
and $^{7}Be$ neutrino spectra in 10 years of 
data taking. Only the effects due to the gas are taken into account.}
\label{fi:spec1}
\end{figure}

However, in spite of its low value,
the MUNU background is still too high for a solar neutrino TPC. 
The minimum volume of such a TPC
should be at least  
200 $m^{3}$ (which corresponds to
a mass of 0.74 $ton$ of $CF_{4}$) for 
a count rate of about 1 solar event/day (half due to $pp$ 
and half to $^{7}Be$ neutrinos, assuming the Standard Solar 
Model neutrino flux and 
a 100 $keV$
threshold on the electron kinetic energy).
Fig. \ref{fi:spec1} shows the reconstructed energy spectrum of the $pp$ and 
$^{7}Be$ neutrinos interacting during 10 years inside a 200 $m^{3}$ TPC 
filled with $CF_{4}$ at 1 $bar$ pressure.
The electron energy resolution used in the simulation 
has been measured with a prototype TPC, 
whereas the angular resolution is given by Monte Carlo. It takes into account 
the multiple scattering of the 
recoiling electron and the diffusion of the ionization electrons along a 
drift distance of 5 $m$ (with the initial 2 $cm$ of the track used to fit 
the direction). The resolution at the $^{7}Be$ peak is in agreement
with the $^{54}Mn$ peak measurement made at 3 $bars$.

We remark that, thanks to the high density of $CF_{4}$, we can 
have a reasonable
target at atmospheric pressure and
room temperature, and we point out that the existing liquid Argon
TPC of the Icarus experiment has already a $\sim$ 200 $m^{3}$ volume. 
With a solar neutrino signal of 1 event/day, assuming a 300 day
data acquisition time, we can have a background of, at most,
$\sim$19 events/day within a $\pi$ $sr$ solid angle
(this way the signal is higher than 4 times 
the backgroud fluctuation).

To prove that this is reachable we need a count rate of, at most,
3 events/week above 100 $keV$ in the MUNU TPC 
running at 1 $bar$ pressure
(for event we mean a single electron in the fiducial volume of the TPC
without any energy deposition larger than 100 $keV$ in the anti-Compton).
Roughly speaking it is an improvement
by 3 orders of magnitude as compared to what we have now in Bugey.
In particular, the two following background sources need a 
detailed study:
\begin{itemize}
\item Radon: it is probably the most subtle background source.
In MUNU, after having replaced the oxisorb filter, we have an
activity of about 0.5 $mBq/m^{3}$. The solar neutrino experiment
is feasible only if the the Radon activity could be decreased
to the $\mu Bq/m^{3}$ level. An activity lower than 0.5 $\mu Bq/m^{3}$
has already been achieved for Nytrogen purified in the liquid
phase\cite{heid}. We have to study if a similar treatment were 
possible for $CF_{4}$.
\item $^{14}C$: it beta decays with the end-point at 156 $keV$
(half-life:5730 years) and it is an extremely severe background source 
for the detection of $pp$ neutrinos. In the atmosphere there are about
10$^{-12}$ $gr$ of $^{14}C$ per $gr$ of $^{12}C$. Borexino has measured
1.8$\cdot$10$^{-18}$ $gr/gr$ of $^{14}C$ in 
liquid scintillator\cite{bor}.
Its content in $CH_{4}$ (used for the production of $CF_{4}$)
in not known. For the $pp$ 
neutrino detection it cannot be higher than 2$\cdot$10$^{-19}$
$gr/gr$. Such a content would produce 10 events/day in the
200 $m^{3}$ solar TPC (within a solid angle of $\pi$ $sr$) and 1.5
events/week in the MUNU TPC. 
\end{itemize}

It is clear that the decisive factor
in a solar neutrino experiment is the background
and that the attainment of the required background 
can be proved only in an underground laboratory.
Because of this I think that the 
next step towards a solar neutrino TPC should be 
the installation of a prototype
in an underground laboratory
as a counting test facility.
\newpage
\section{Summary}
I described the two methods which are studied for the spectroscopy 
of the $pp$ and $^{7}Be$ neutrinos from the Sun:
absorption in a loaded liquid scintillator and elastic scattering 
in a TPC.

From the point of view
of the physics the two approaches are complementary, since the 
charged current interaction is only sensitive to $\nu_{e}$ whereas
the neutral current one is also sensitive to $\nu_{\mu}$ and
$\nu_{\tau}$. 

Background suppression is probably the most difficult task 
for both methods. I think that the richer information
provided by a TPC might be a decisive factor to achieve the required
background. On the other hand a liquid scintillator detector 
is clearly a much 'easier' detector than a TPC.  

The capability of suppressing the background
to the required level 
can only be
proven for both methods by a prototype measuring in an
underground laboratory.
\section{Acknowledgments}
I wish to tank my colleagues of the MUNU experiment for all the work
done together, M. Baldo Ceolin and G. Costa for the organization
of a stimulating workshop in a unique atmosphere.


\begin{thebibliography}{99}
\bibitem{ba2000} J.N. Bahcall, {\it Nucl. Phys.} {\bf B100} (2001) 5.
\bibitem{ba2001} J.N. Bahcall, {\it hep-ex} {\bf 0106086v1} (2001).
\bibitem{ra97} R.S. Raghavan, {\it Phys. Rev. Lett.} {\bf 78} (1997) 3618.
\bibitem{ra01} R.S. Raghavan, {\it Rep. to LNGS Scient. Comm.} (3-2001).
\bibitem{ra02} R.S. Raghavan, {\it hep-ex} {\bf 0106054} (2001).
\bibitem{hel01} J. Seguinot et al., {\it LPC 92-31hep-ex} (1992).
\bibitem{bon01} G. Bonvicini et al., {\it hep-ex} {\bf 0109032} (2001).
\bibitem{bro92} C. Broggini et al., {\it NIM} {\bf A311} {1992} 319. 
\bibitem{am97} C. Amsler et al., {\it NIM} {\bf A396} {1997} 115. 
\bibitem{av01} M. Avenier et al., {\it hep-ex} {\bf 0106104v1} (2001).
\bibitem{bro93} C. Broggini , {\it NIM} {\bf A332} {1993} 413. 
\bibitem{heid} G. Heusser et al., {\it Appl. Rad. and Isot.} {\bf 53} {2000} 371. 
\bibitem{bor} G. Alimonti et al., {\it Astr.Phys.} {\bf 8} {1998} 141.
\end{thebibliography}
\end{document}